\documentstyle[12pt,epsfig,axodraw]{article}
\def\be{\begin{equation}}
\def\ee{\end{equation}}
\def\bc{\begin{center}}
\def\ec{\end{center}}
\def\bea{\begin{eqnarray}}
\def\eea{\end{eqnarray}}

\def\ph{\varphi}
\def\phb{{\ov\varphi}}
\def\hphi{{\hat{\phi}}}
\def\hphib{{\hat{{\ov\phi}}}}
\def\hph{{\hat{\varphi}}}
\def\hphb{{\hat{\ov\varphi}}}
\def\hv{\hat V}
\def\hf{{\hat{F}}}
\def\hfb{{\hat{\ov F}}}
\def\hd{\hat{D}} 
\def\th{\theta}
\def\thb{\ov\theta}
\def\psib{\ov\psi}
\def\fb{\ov{F}}
\def\chib{\ov\chi}
\def\gb{\ov{G}}
\def\la{\lambda}
\def\lab{\ov\lambda}
\def\hk{\hat{K}}
\def\hx{\hat{S}}
\def\hr{\hat{H}}
\def\hw{\hat{w}}
\def\hwb{\hat{\ov w}}
\def\hff{\hat{f}}
\def\cl{{\cal L}}
\def\cm{{\cal M}}
\def\co{{\cal O}}
\def\dmu{\partial^\mu}
\def\dmd{\partial_\mu}
\def\dnu{\partial^\nu}
\def\dnd{\partial_\nu}

\def\lsq{{\Lambda^2}}
\def\ov{\overline}

\def\smu{\sigma^{\mu}}

\def\smub{{\ov\sigma}^{\mu}}

\begin{document}
\begin{titlepage}
\vspace*{-1cm}
\phantom{hep-th/0001121} 
\hfill{DFPD-00/TH/03}
\vskip 2.0cm
\begin{center}
{\Large\bf One-loop K\"ahler potential} 
\vskip 0.3 cm
{\Large\bf in non-renormalizable theories} 
\end{center}
\vskip 1.5  cm
\begin{center}
{\large 
Andrea Brignole\footnote{E-mail address: 
 brignole@padova.infn.it}
\\
\vskip .5cm
Istituto Nazionale di Fisica Nucleare, Sezione di Padova, 
\\
Dipartimento di Fisica `G.~Galilei', Universit\`a di Padova, 
\\
Via Marzolo n.8, I-35131 Padua, Italy}
\end{center}
\vskip 3.0cm
\begin{abstract}
\noindent

We consider a general $d$=4 $N$=1 globally supersymmetric 
lagrangian involving chiral and vector superfields, 
with arbitrary superpotential, K\"ahler potential
and gauge kinetic function. We compute perturbative quantum 
corrections by employing a component field approach that 
respects supersymmetry and background gauge invariance.
In particular, we obtain the full one-loop correction to 
the K\"ahler potential in supersymmetric Landau gauge.
Two derivations of this result are described.
The non-renormalization of the superpotential and the 
quadratic correction to the Fayet-Iliopoulos terms are 
further checks of our computations. 

\end{abstract}
\end{titlepage}
\setcounter{footnote}{0}
\vskip2truecm

\section{Introduction}

Perturbative quantum corrections have a peculiar form in 
supersymmetric theories. In the case of $d$=4 $N$=1 theories, in 
particular, the non-renormalization theorem \cite{grs} establishes 
that the superpotential remains uncorrected, whereas the K\"ahler 
potential generically receives quantum corrections.
The anomalous dimensions of chiral superfields are the simplest 
example of the latter effects, but a richer structure emerges 
if the full field dependence of such corrections is taken into account. 
Full one-loop corrections to the K\"ahler potential have
been recently computed, both in the Wess-Zumino model \cite{bky} 
and in more general renormalizable models \cite{wgr,pw,gru}.
For the most general renormalizable $N$=1 theory, the one-loop
correction to the K\"ahler potential was found to have
a very compact form in supersymmetric Landau gauge.
The result reads \cite{gru}:
\be
\label{kren}
\Delta K =
- \, {1 \over 32 \pi^2} \left[
{\rm Tr} \left( 
\cm^2_{\phi} \left( \log {\cm^2_{\phi} \over \lsq} \, -1 \right) \right)
- 2 \, {\rm Tr} \left(
\cm^2_V \left( \log {\cm^2_V \over \lsq} \, -1 \right) \right)
\right]
\ee
where $\cm^2_{\phi}$ and $\cm^2_V$ are the (chiral superfield 
dependent) mass matrices in the chiral and vector superfield sectors,
respectively, and $\Lambda$ is an ultraviolet cutoff.
One-loop corrections to the K\"ahler potential have also been 
investigated in non-renormalizable $N$=1 models, divergent 
contributions being the main focus. For instance,  
quadratically divergent corrections to the K\"ahler 
potential in general models were computed in \cite{grk}.
Quadratic and logarithmic divergences were also studied  
in general supergravity models with diagonal gauge kinetic 
function \cite{maryk}, or in models with chiral superfields only 
\cite{bpr}. In \cite{cl}, the Wilsonian evolution of the K\"ahler 
potential was studied in non-renormalizable models with an abelian 
vector superfield and/or gauge singlet chiral superfields.
Divergent and finite corrections in specific models were
also evaluated in \cite{bcp}.

The main purpose of this paper is to generalize the result 
(\ref{kren}) to non-renormalizable theories, i.e. to compute
the full (divergent and finite) one-loop correction to the 
K\"ahler potential in a general globally supersymmetric 
theory. Our perturbative calculation starts from
a tree-level lagrangian in which the superpotential, the 
K\"ahler potential, the gauge kinetic function, the gauge
group and the matter representations are arbitrary. 
Upon quantizing the theory, a supersymmetric gauge fixing 
term is added, and we choose to preserve supersymmetric 
background gauge invariance. 
This framework is then translated to the component field level
and quantum corrections are computed in terms of component 
Feynman diagrams. Notice that we
do {\it not} choose the Wess-Zumino gauge, supplemented by
a gauge fixing term for the component vector fields.
Instead, we keep all components of quantum supermultiplets 
and use supersymmetric Landau gauge. Thus our component 
computations are equivalent to superfield computations, and 
a superfield language can be used to interpret our results. 
In particular, we obtain the full one-loop correction to the
K\"ahler potential, which is our main result. In spite of 
the fact that the interactions are considerably more 
complicated in comparison to the renormalizable case, we find 
that the logarithmically divergent and finite one-loop corrections 
to the K\"ahler potential can be cast in the same form as in 
eq.~(\ref{kren}), with generalized mass matrices $\cm^2_{\phi}$ 
and $\cm^2_V$. In addition to that, the K\"ahler potential receives
a quadratically divergent correction, in agreement with~\cite{grk}.
A consistency check based on supersymmetric background gauge 
invariance is also discussed.
The non-renormalization of the superpotential and the quadratic 
correction to the Fayet-Iliopoulos terms are further checks of
our computations.

\section{Theoretical framework}

We consider a general $d$=4 \ $N$=1 globally supersymmetric theory defined 
by a tree-level lagrangian of the form (see e.g. \cite{wb,ggrs}):
\be
\label{lclas}
{\cal L} \! = \! 
\left[ \int \!\! d^2 \th \, w( \hphi ) +  {\rm h.c.} \right]
+ \int \!\! d^4 \th \left[ K (\hphib, e^{2\hv} \hphi )
+ 2 \kappa_a \hv^a \right]
+ \left[ \int \!\! d^2 \th \,{1 \over 4} f_{ab}( \hphi ) 
{\hat{\cal W}}^a {\hat{\cal W}}^b +  {\rm h.c.} \right] .
\ee
The theory has a general (possibly product) gauge group $G$,
with hermitian generators $T_a$ satisfying the
Lie algebra $[T_a,T_b]=i c_{ab}^{\;\;\;c} T_c$. The associated
vector superfields $\hv=\hv^a T_a$ have superfield 
strengths ${\hat{\cal W}}= {\hat{\cal W}}^a T_a=
-{1 \over 8} {\ov D}_{\dot{\alpha}} {\ov D}^{\dot{\alpha}} e^{-2\hv} 
D_{\alpha} e^{2 \hv}$. The chiral superfields $\hphi=\{\hphi^i\}$ 
belong to a general (reducible) representation of $G$.
Supersymmetric gauge transformations read
$e^{2\hv} \rightarrow e^{-2 i \hat\Lambda^{\dagger}} 
e^{2\hv} e^{2 i \hat\Lambda}$ and $\hphi \rightarrow  
e^{-2 i \hat\Lambda} \hphi$, where $\hat\Lambda = \hat\Lambda^a T_a$
is chiral.
The Fayet-Iliopoulos coefficients $\kappa_a$ are real and may be 
nonvanishing only for the abelian factors of the gauge group $G$. 
The superpotential $w$, the K\"ahler potential $K$ and the gauge
kinetic function $f_{ab}$ are only constrained by gauge invariance,
and are otherwise arbitrary. In more detail:
$w(\hphi)$ is $G$-invariant, $K({\hphib},\hphi)$ is real and 
$G$-invariant, and $f_{ab}(\hphi)$ transforms as a symmetric 
product of adjoint representations of $G$. These constraints are 
expressed by the identities:
\bea
\label{wid}
w_i(\hphi)(T_a \hphi)^i \! & \equiv &\! 0
\\
\label{kid}
K_i(\hphib,\hphi)(T_a \hphi)^i  \!  & \equiv & \! 
(\hphib T_a)^{\ov\imath} K_{\ov\imath}(\hphib,\hphi) 
\\
\label{fid}
f_{ab \, i}(\hphi)(T_c \hphi)^i  \!  & \equiv & \! 
i c_{ac}^{\;\;\;d} f_{db}(\hphi) + i c_{bc}^{\;\;\;d} f_{ad}(\hphi) 
\eea
where $(T_a \hphi)^i \equiv (T_a)^i_{\, j}\hphi^j$,  
$(\hphib T_a)^{\ov\imath} \equiv \hphib\,^{\ov\jmath}
(T_a)_{\ov\jmath}^{\;\; \ov\imath}\equiv 
({\ov T}_a)^{\ov\imath}_{\, \ov\jmath}\hphib\,^{\ov\jmath}$,
$w_i(\hphi)\equiv \partial w(\hphi)/\partial\hphi^i$, and so on.
Further identities can be obtained by differentiating the ones 
above. We also recall that, in the special case of a renormalizable 
theory, $w$ is at most cubic in the chiral superfields, $K$ is 
quadratic (i.e. canonical) and $f_{ab}$ is constant (i.e. canonical). 
Here we aim at full generality and do not impose renormalizability.

The supersymmetric lagrangian above is the most general one that 
contains no more than two space-time derivatives on component 
fields\footnote{This is already a non-trivial generalization 
of the renormalizable case. A further generalization would be the 
inclusion of supersymmetric higher derivative terms, which are
additional non-renormalizable terms. We leave this to
future investigation.}. It could arise as a low-energy limit of 
a more fundamental theory. In fact,
since we take $w,K,f_{ab}$ to be generic functions, not restricted
by renormalizability, the lagrangian $\cl$ in eq.~(\ref{lclas}) 
necessarily describes an effective theory, valid below some cutoff 
scale $\Lambda$. For consistency reasons, $\Lambda$
cannot be much larger than the scale whose inverse powers 
control the non-renormalizable terms in $w,K,f_{ab}$. 
Anyhow, we will not address the origin of the lagrangian itself. 
We will just take $\cl$ as a general classical 
bare lagrangian and study the corresponding one-loop
corrections. In principle, such self-corrections
could be matched to those of the hypothetical underlying 
theory, if the latter theory were known.

Quantum corrections to a classical supersymmetric lagrangian
can be computed by various techniques. We find it convenient
to use the background field method \cite{grs,ggrs}. 
The superfields $\hphi,\hv$ are split into background (still 
denoted by $\hphi,\hv$) and quantum ($\phi,V$) parts.
A supersymmetric gauge fixing lagrangian $\cl_{\rm gf}$ is added 
in order to break the quantum gauge invariance, and a corresponding 
ghost lagrangian $\cl_{\rm gh}$ is introduced. 
If we choose to preserve supersymmetric background gauge invariance, 
we can take \cite{grs,ggrs}:
\bea
\label{phsplit}
\hphi & \longrightarrow & \hphi + \phi \, ,
\\
\label{vsplit}
e^{2 \hv} & \longrightarrow & e^{\hv} e^{2 V} e^{\hv} \, ,
\eea
\be 
\label{gfix}
\cl_{\rm g f} =  -{1 \over 8 \xi} \int \! d^4 \theta
\, (\nabla^2 V)^a \, ({\ov \nabla}^2 V)^a \, ,
\ee
where the background vector superfield has been put in a convenient
form, $\xi$ is a gauge parameter 
and $\nabla_{\alpha} \equiv e^{-\hv} D_{\alpha} e^{\hv}$, 
${\ov \nabla}_{\dot{\alpha}} \equiv e^{\hv} {\ov D}_{\dot{\alpha}} 
e^{-\hv}$ are background gauge covariant supersymmetric derivatives.
Ghost superfields interact with vector superfields only, and
we will not need the explicit expression of $\cl_{\rm gh}$. 
In the abelian case, the splitting of the vector superfield 
in (\ref{vsplit}) reduces to $\hv \longrightarrow \hv + V$ 
and the gauge fixing term (\ref{gfix}) becomes
\be
\label{abel}
\cl_{\rm g f} = -{1 \over 8 \xi} \int \! d^4 \theta
\, D^2 V^a \, {\ov D}^2 V^a \, .
\ee
Making these simpler choices in the non-abelian case is certainly 
allowed, but does not lead to a background gauge invariant 
effective action. Since we find it useful to preserve the latter
property, we proceed in the way explained above. 

Once the replacements (\ref{phsplit}), (\ref{vsplit})
have been made in (\ref{lclas}) and the gauge
fixing and ghost terms have been added, the
resulting lagrangian can be expanded in powers of
the quantum superfields.   
The zero-th order part is just the original lagrangian
(\ref{lclas}) for the classical (background) 
superfields, $\cl(\hphi,\hv)$. The terms linear in 
quantum superfields do not contribute to the 
(one-particle-irreducible) effective action and can
be dropped. The part bilinear in quantum superfields, 
$\cl_{\rm bil}(\phi,V,\ldots;\hphi,\hv)$, is the relevant one for 
the computation of one-loop quantum corrections. 
One has to integrate out the quantum superfields  
in the theory defined by $\cl+\cl_{\rm bil}$, either diagrammatically 
or by direct functional methods. The result of this 
operation gives the one-loop-corrected effective action,
a functional of background superfields only. 
If a (super) derivative expansion of the latter
functional is performed, the lowest order terms can 
be interpreted as corrections to the basic functions
in (\ref{lclas}). In other words, one-loop 
corrections to the effective lagrangian will have 
the form:
\be
\label{deltal}
\Delta \cl =   
\left[ \int \! d^2 \theta \, \Delta w( \hphi ) + \, {\rm h.c.} \right]
+ \int \! d^4 \theta \, \left[ \Delta K (\hphib, e^{2\hv} \hphi )
+ 2 \, \Delta\kappa_a \hv^a \right] \, + \ldots
\ee
where the dots stand for terms containing supercovariant
derivatives (${\hat{\cal W}} {\hat{\cal W}}$ terms and
higher derivative terms), which we will not compute.
Our purpose is to compute the corrections $\Delta w$,
$\Delta K$ and $\Delta \kappa_a$. 

\section{Component field approach}

The expected form (\ref{deltal}) of  quantum corrections
relies on the assumption that quantum superfields are integrated
out in a supersymmetric way. This is automatic if the
perturbative computations are performed at the superfield level.
Here we choose to work with component fields, instead. 
Nevertheless, we retain the whole off-shell structure of quantum 
supermultiplets and literally translate the framework described 
above to the component level. In particular, instead of simplifying 
the structure of vector supermultiplets by using the Wess-Zumino gauge, 
supplemented by a gauge fixing term for the component vector fields 
only, we work with full supermultiplets and use the manifestly 
supersymmetric gauge fixing term (\ref{gfix}). 
Thus integrating out component fields is literally equivalent 
to integrating out superfields, and supersymmetric background
gauge invariance can also be preserved.

In order to fix the notation, we recall the component expansion of 
chiral and vector superfields\footnote{Our conventions 
are slightly different from those of ref.~\cite{wb}. For instance,
we use the space-time metric $g_{\mu\nu}={\rm diag}(+1,-1,-1,-1)$, 
the Pauli matrices $\smu=(1,\vec{\sigma})$, $\smub=(1,-\vec{\sigma})$, 
and the supersymmetric derivatives $D_{\alpha}=\partial/\partial 
\th^{\alpha} - i \sigma^{\mu}_{\alpha{\dot{\alpha}}} 
\thb^{\dot{\alpha}} \dmd$, ${\ov D}_{\dot\alpha}=-\partial/\partial 
\thb^{\dot\alpha} + i \th^{\alpha} \sigma^{\mu}_{\alpha{\dot{\alpha}}} 
\dmd$.}:
\bea
\phi^i &  = &  \ph^i + \sqrt{2} \th \psi^i + \th \th F^i 
- i \th \smu \thb \dmd \ph^i
- {i \over \sqrt{2}} \th \th \thb \smub \dmd \psi^i
- {1 \over 4} \th \th \thb \thb \Box \ph^i
\\
V^a & = & C^a +  i \th \chi^a - i \thb \chib^a 
+ {i \over \sqrt{2}} \th \th G^a - {i \over \sqrt{2}} 
\thb \thb \gb^a + \th \smu \thb A_{\mu}^a
\nonumber \\ & & 
+ i \th\th \thb \left(\lab^a - {i \over 2} \smub \dmd \chi^a\right) 
- i \thb\thb \th \left(\la^a - {i \over 2} \smu \dmd \chib^a\right) 
+{1 \over 2} \th \th \thb \thb \left(D^a - {1\over 2}\Box C^a \right)
\eea
where $C^a,D^a$ are real scalar fields, $\ph^i,F^i,G^a$ are complex
scalar fields, $A_{\mu}^a$ are real vector fields, and
$\psi^i,\chi^a,\la^a$ are complex Weyl fields.
The expansions above apply to the quantum superfields $\phi^i$
and $V^a$. Similar component expansions hold for the background 
superfields $\hphi^i$ and $\hv^a$, and one can eventually obtain
the full component expansion of $\cl_{\rm bil}$,
to be used for the computation of $\Delta \cl$. However,
since supersymmetry constrains one-loop corrections to have the 
form (\ref{deltal}), a convenient choice of the background superfields 
can simplify the computation of the functions $\Delta K$ and $\Delta w$.
For instance, in order to compute the function $\Delta K$, we 
could choose a background in which the only non-vanishing fields 
are the scalars $\hph^i(x)$ contained in $\hphi^i$.
Then eq.~(\ref{deltal}) predicts that the one-loop computation 
should produce terms of the form $\Delta K_{{\ov\imath} j}(\hphb,\hph) 
\dmd\hphb\,^{\ov\imath} \dmu\hph^j$. Hence the one-loop correction 
$\Delta K_{{\ov\imath} j} (\hphb,\hph)$ to the K\"ahler metric could 
be identified and the functional form of the one-loop-corrected
K\"ahler potential could be reconstructed by integration.
Alternative choices of the background can be even more convenient.
In what follows, we will make this choice:
\bea
\label{backp}
\hphi^i & = & \hph^i + \th\th \, \hf^i \, ,
\\
\label{backv}
\hv^a & = &  {1\over 2} \, \th\th \thb\thb \, \hd^a \, ,
\eea
where both the physical scalars $\hph^i$ and the auxiliary fields $\hf^i,
\hd^a$ are taken to be constant (i.e. space-time independent).
If we specialize (\ref{deltal}) to this background, we infer that
the one-loop computation should generate the terms:
\be
\label{deltall}
\Delta \cl = \left[ \Delta w_i(\hph) \hf^i + {\rm h.c.} \right] 
+ \Delta K_{{\ov\imath} j}(\hphb,\hph) \hfb\,^{\ov\imath} \hf^j  
+ \left[ \Delta K_j (\hphb,\hph) (T_a \hph)^j 
+ \Delta \kappa_a \right] \hd^a + \ldots
\ee
The dots stand for higher order terms in $\hf,\hd$ and
correspond to the (omitted) ${\hat{\cal W}} {\hat{\cal W}}$ and
higher derivative terms in (\ref{deltal}). 
The computational advantage of using a background with constant 
$\hph$, $\hf$, $\hd$ is obvious: the one-loop diagrams to be 
evaluated have vanishing external momenta. In other words,  
in such a background $-\Delta \cl$ is the one-loop correction to the 
effective potential ${\cal V}_{\rm eff}(\hph,\hf,\hd)$, considered 
as a function of both physical and auxiliary scalar fields.
In the case of renormalizable models, one-loop computations
of this object by either component or superfield techniques can be
found, for instance, in refs.~\cite{effp,effps}.
Here we are interested in obtaining the first few terms in the 
auxiliary field expansion of the effective potential, in the 
general non-renormalizable theory defined above.  
Once the results of the one-loop computation have been cast
in the form (\ref{deltall}), the functions $\Delta w$ and 
$\Delta K$ can be reconstructed from their derivatives\footnote{
Up to irrelevant constant terms in  
$\Delta w$ or harmonic terms in $\Delta K$.} $\Delta w_i$,
$\Delta K_j,\Delta K_{{\ov\imath} j}$, and the coefficients 
$\Delta \kappa_a$ can be easily identified, too. Notice that the 
one-loop correction to the K\"ahler potential can be reconstructed 
in two different ways, thanks to background gauge invariance: 
this allows us to make a non-trivial consistency check. 
In fact, we will proceed as follows. First, we will expand 
the lagrangian in a background with constant $\hph,\hf$ and
vanishing $\hd$ (section~4). Then we will compute the one-loop 
corrections to the terms linear and 
quadratic in $\hf$, from which $\Delta w$ and $\Delta K$ can be
reconstructed (sections~5 and 6). In the final part 
(section~7) we will instead consider a background with 
constant $\hph,\hd$ and vanishing $\hf$. Then we will compute 
the one-loop corrections to the terms linear in $\hd$. This will allow 
us to identify $\Delta \kappa_a$ and to derive $\Delta K$ in a different 
way.
 
\section{Quantum bilinears and propagators}

We take the background chiral superfields in the form 
(\ref{backp}), with constant $\hph,\hf$ fields, and
plug the background-quantum splitting (\ref{phsplit})
in (\ref{lclas}).
At the same time, we take vanishing background vector
superfields\footnote{Background gauge invariance will not be 
exploited directly till section~7.}. In this case, the background-quantum
splitting (\ref{vsplit}) reduces to the simple replacement 
$\hv \rightarrow V$ in (\ref{lclas}), $\cl_{\rm gf}$ in 
(\ref{gfix}) reduces to (\ref{abel}), 
and $\cl_{\rm gh}$ is not relevant because ghosts do not interact
with the (chiral) background. The lagrangian is then 
expanded and only the part bilinear in the quantum fields $\phi,V$
is retained, $\cl_{\rm bil}=\cl_{\phi\phi} +\cl_{VV}+\cl_{\phi V}$.
Integration by parts is used whenever convenient, and
some relations that follow from (\ref{kid}) are used to 
rearrange the terms generated by the expansion of the 
K\"ahler potential.
The terms that make up $\cl_{\rm bil}$ have a complicated dependence 
on $\hph$ and a simple dependence (at most quadratic) on $\hf$.
Thus each of the three parts in $\cl_{\rm bil}$ can in turn be
decomposed as $\cl_{\phi\phi}=\cl^{(0)}_{\phi\phi}+
\cl^{(\hf)}_{\phi\phi}+\cl^{(\hf\hf)}_{\phi\phi}$, 
$\cl_{VV}=\cl^{(0)}_{VV}+\cl^{(\hf)}_{VV}+\cl^{(\hf\hf)}_{VV}$, 
$\cl_{\phi V}=\cl^{(0)}_{\phi V}+
\cl^{(\hf)}_{\phi V}+\cl^{(\hf\hf)}_{\phi V}$, in a 
self-explanatory notation.

First of all we list the terms that do not depend on $\hf$, i.e.
the quantum bilinears in a pure constant $\hph$ background:
\bea
\label{lpp}
\cl^{(0)}_{\phi\phi} & = & 
\hk_{\ov\imath j} \left(- \phb^{\ov\imath} \Box \ph^j
+ \fb^{\ov\imath} F^j + i \psib^{\ov\imath} \smub \dmd \psi^j \right)
+ \left[ \hw_{ij} \left( F^i \ph^j 
- {1\over 2} \psi^i \psi^j \right) + \, {\rm h.c.} \right]
\\
\label{lvv}
\cl^{(0)}_{V V} & = &\!\! \hr_{ab} \left[
- {1\over 4} (\dmd A_{\nu}^a - \dnd A_{\mu}^a)
(\dmu A^{\nu b} - \dnu A^{\mu b})
+ {1\over 2} D^a D^b + i \lab^a \smub \dmd \la^b \right]
\nonumber\\ & + &
\!\! \hx_{ab}
\left[ {1\over 2} A_{\mu}^a A^{\mu b} 
+ C^a D^b - {1\over 2} C^a \Box C^b + \gb^a G^b
-\left( \chi^a \la^b - {i \over 2} \chi^a \smu \dmd \chib^b 
+ {\rm h.c.} \right) \right]
\nonumber\\ & - & 
\! \! { 1  \over \xi}  \left[ 
{1\over 2}(\dmu A_{\mu}^a)^2 \! +\! {1\over 2}(D^a\!-\! \Box C^a)^2 
\!+\! \dmu \gb^a \dmd G^a 
\!+\!  i (\lab^a\! +\! i \partial_{\rho} \chi^a \sigma^{\rho} ) 
{\ov\sigma}^{\nu} \dnd  (\la^a\! -\! i \smu \dmd \chib^a) \right] 
\nonumber\\ & &
\\
\label{lpv}
\cl^{(0)}_{\phi V} & = &
(\hphb T_a)^{\ov\imath} \hk_{\ov\imath j} 
\left[ \ph^j (i \dmu A^a_{\mu}+ D^a - \Box C^a) 
- i\sqrt{2} F^j \gb^a 
+ i\sqrt{2} \psi^j (\la^a - i \smu \dmd \chib^a)
\right] + {\rm h.c.} 
\nonumber\\ & &
\eea
Most of the dependence on $\hph$ is left implicit. In particular, 
we have used the abbreviations:
\bea
\label{hkhw}
\hk_{\ov\imath j}\equiv K_{\ov\imath j}(\hphb,\hph)
\, , \phantom{xxxxx}
& &  
\;\;\;\;\;\;\;\;\;
\hw_{ij} \equiv w_{ij}(\hph) \, , 
\\
\label{hrhx}
\hr_{ab} \equiv {1 \over 2}
\left(\! f_{ab}(\hph) \!+\! {\ov f}_{ab}(\hphb) \! \right) 
\, , & & 
\!\!\!\!\!\! \hx_{ab} \equiv 
(\hphb T_a)^{\ov\imath} K_{\ov\imath j}(\hphb,\hph) (T_b \hph)^j 
\!+\!(\hphb T_b)^{\ov\imath} K_{\ov\imath j}(\hphb,\hph) (T_a \hph)^j .
\eea
The matrices $\hk_{\ov\imath j}$ and $\hw_{ij}$ ($\hr_{ab}$ 
and $\hx_{ab}$) have the meaning of $\hph$ dependent 
metric and masses in the $\phi$ ($V$) sector. Notice
that $\hk_{\ov\imath j}$ is hermitian and $\hw_{ij}$ is symmetric, 
whereas $\hr_{ab}$ and $\hx_{ab}$ are real and symmetric.
Next we list the terms that depend on $\hf$:
\bea
\label{lppf}
\cl^{(\hf)}_{\phi\phi} & = & 
\hf^j \left[ {1 \over 2} \hw_{i k j} \ph^i \ph^k
+ \hk_{ \ov\imath k j} \fb^{\ov\imath} \ph^k
+ \hk_{ \ov\imath \ov{k} j} 
\left( \fb^{\ov\imath} \phb^{\ov{k}} 
- { 1\over 2} \psib^{\ov\imath} \psib^{\ov{k}} \right)
\right] + {\rm h.c.}
\\
\label{lvvf}
\cl^{(\hf)}_{V V} & = & 
\hf^j \left[ 
- {1 \over 2} (\hr_{ab})_j \la^a \la^b
+ (\hx_{ab})_j \left( {1 \over 2} \chib^a \chib^b
-i \sqrt{2} C^a \gb^b \right)
\right] + {\rm h.c.}
\\
\label{lpvf}
\cl^{(\hf)}_{\phi V} & = & 
\hf^j \left[ 
\left( \hk_{\ov\imath k} (T_a)^k_{\; j}
+ \hk_{\ov\imath k j} (T_a \hph)^k \right)
\left( 2 \fb^{\ov\imath} C^a  
+ i \sqrt{2} \psib^{\ov\imath} \chib^a
- i \sqrt{2} \phb^{\ov\imath} \gb^a \right) \right.
\nonumber\\
& & 
\phantom{\hf^j}  \left. 
- i \sqrt{2} (\hphb T_a)^{\ov\imath}
\hk_{\ov\imath k j} \ph^k \gb^a 
\right] + {\rm h.c.}
\\
\label{lppff}
\cl^{(\hf\hf)}_{\phi\phi} & = & 
\hfb\,^{\ov\imath} \hf^j \left[
\hk_{\ov\imath j \ov{k} \ell} \phb^{\ov{k}} \ph^{\ell}
+ {1 \over 2} \left( \hk_{\ov\imath j k \ell} 
\ph^k \ph^{\ell} + \hk_{\ov\imath j \ov{k} \ov\ell} 
\phb^{\ov{k}} \phb^{\ov\ell} \right)
\right]
\\
\label{lvvff}
\cl^{(\hf\hf)}_{V V} & = & 
\hfb\,^{\ov\imath} \hf^j (\hx_{ab})_{\ov\imath j} C^a C^b
\\
\label{lpvff}
\cl^{(\hf\hf)}_{\phi V} & = & 
\hfb\,^{\ov\imath} \hf^j 
\left[ 2 \left( (T_a)_{\ov\imath}^{\; \ov\ell} \hk_{\ov\ell k j}
+(\hphb T_a)^{\ov\ell} \hk_{\ov\ell k \ov\imath j} \right) 
\ph^k C^a \right] + {\rm h.c.}
\eea

Having completed the list of quantum bilinears in the chosen
background, we could in principle compute the one-loop effective 
potential ${\cal V}_{\rm eff}(\hph,\hf,\hd\!=\!0)$, which corresponds 
to diagrams with an arbitrary number of (zero momentum) $\hph$ 
and $\hf$ external legs. In practice, as explained above,
we are only interested in the first few terms in the $\hf$
expansion of the effective potential. This amounts to compute 
diagrams with an arbitrary number of $\hph$ external legs and a 
small number of $\hf$ external legs. In order to take 
into account the full $\hph$ dependence, we will proceed as follows.
In the remainder of this section, we will consider the quantum 
bilinears in the pure constant $\hph$ background and compute the 
$\hph$-dressed propagators for the quantum fields. 
In the next two sections, we will consider 
the quantum bilinears that also depend on $\hf$ and treat 
them as $\hph$- and $\hf$-dependent interaction vertices, to be 
joined by the $\hph$-dressed quantum propagators. In particular, 
we will compute the $\hph$-dressed one- and two-point functions 
of $\hf$, which will allow us to reconstruct $\Delta w$ and $\Delta K$,
respectively.

The quantum propagators in a constant $\hph$ background 
are obtained by inverting the quadratic forms that appear
in (\ref{lpp}), (\ref{lvv}), (\ref{lpv}). We omit
the details of the derivation and write the results directly 
in momentum space. We use a compact matrix notation and
denote by $\hk,\hw,\hr,\hx$ the $\hph$ dependent matrices 
defined in (\ref{hkhw}) and (\ref{hrhx}).
The matrices $\hk$ and $\hw$ should 
not be confused with the functions $K$ and $w$ used to define 
them by double differentiation (the context should make this
distinction clear, despite the slight abuse of notation).

From $\cl^{(0)}_{\phi\phi}$, eq.~(\ref{lpp}), we obtain the 
$\hph$-dressed propagators for the components of quantum chiral 
superfields:
\bea
\label{cprop}
< \ph^i \, \phb^{\ov \jmath} >_{\hph}
& = & i \left[ 
\left( \hk p^2 - \hwb \hk^{-1 \; T} \hw \right)^{-1}
\right]^{i \ov\jmath}
\\
< F^i \, \fb^{\ov \jmath} >_{\hph} 
& = & i p^2 \left[ 
\left( \hk p^2 - \hwb \hk^{-1 \; T} \hw \right)^{-1}
\right]^{i \ov\jmath}
\\
< F^i \, \ph^j >_{\hph}
& = & - i \left[ 
\left( \hk p^2 - \hwb \hk^{-1 \; T} \hw \right)^{-1}
\hwb \hk^{-1 \; T}  \right]^{i j}
\\ 
< \psi^i_{\alpha} \, \psib^{\ov \jmath}_{\dot{\alpha}}>_{\hph} 
& = & i p_{\mu}\smu_{\alpha \dot{\alpha}}
\left[ 
\left( \hk p^2 - \hwb \hk^{-1 \; T} \hw \right)^{-1}
\right]^{i \ov\jmath}
\\
< \psi^i_{\alpha} \, \psi^{\beta j} >_{\hph} 
& = &  i \delta_{\alpha}^{\; \beta}
\left[ 
\left( \hk p^2 - \hwb \hk^{-1 \; T} \hw \right)^{-1}
\hwb \hk^{-1 \; T} \right]^{i j}
\eea

From $\cl^{(0)}_{V V}$, eq.~(\ref{lvv}), we obtain the 
$\hph$-dressed propagators for the components of quantum 
vector superfields:
\bea
\label{vprop}
< A_{\mu}^a \,  A_{\nu}^b >_{\hph} 
& = & 
i \left( -g_{\mu \nu} + {p_{\mu} p_{\nu} \over p^2} \right)
\left[ \left(\hr \, p^2 - \hx \right)^{-1} \right]^{a b} 
-  i \xi {p_{\mu} p_{\nu} \over p^2} 
\left[ \left( p^2 - \xi \hx \right)^{-1} \right]^{a b} 
\\
< G^a \, \gb^b >_{\hph} 
& = & 
- i \xi \left[ \left( p^2 - \xi \hx \right)^{-1} \right]^{a b} 
\\
< D^a \, D^b >_{\hph} 
& = & 
i p^2  \left[ \left(\hr \, p^2 - \hx \right)^{-1} \right]^{a b}
\\ 
< D^a \, C^b >_{\hph} 
& = & 
- i \left[ \left(\hr \, p^2 - \hx \right)^{-1} \right]^{a b}
\\ 
< C^a \, C^b >_{\hph} 
& = &  i {1 \over  p^2} 
\left[ \left(\hr \, p^2 - \hx \right)^{-1} \right]^{a b}
- i \xi {1 \over  p^2} 
\left[ \left( p^2 - \xi \hx \right)^{-1} \right]^{a b}
\\
< \la^a_{\alpha} \, \lab^b_{\dot{\alpha}} >_{\hph} 
& = & 
i p_{\mu}\smu_{\alpha \dot{\alpha}}
 \left[ \left(\hr \, p^2 - \hx \right)^{-1} \right]^{a b}
\\ 
< \la^a_{\alpha} \, \chi^{\beta b} >_{\hph} 
& = & 
i \delta_{\alpha}^{\; \beta}
\left[ \left(\hr \, p^2 - \hx \right)^{-1} \right]^{a b} 
\\ 
< \chi^a_{\alpha} \, \chib^b_{\dot{\alpha}} >_{\hph}
& = & 
i { p_{\mu}\smu_{\alpha \dot{\alpha}} \over  p^2} 
\left[ \left(\hr \,  p^2 - \hx \right)^{-1} \right]^{a b}
- i \xi {p_{\mu}\smu_{\alpha \dot{\alpha}} \over  p^2} 
\left[ \left( p^2 - \xi \hx \right)^{-1} \right]^{a b}
\eea

We have not yet taken into account the terms in 
$\cl^{(0)}_{\phi V}$, eq.~(\ref{lpv}), which mix the components 
of chiral and vector superfields. Such terms could be treated as
insertions, to be eventually resummed. 
This task becomes very easy if supersymmetric Landau gauge is 
used, that is, if the special value $\xi=0$ of the gauge 
parameter is chosen.  It is well known that ordinary Landau gauge 
simplifies both the computation and the form of the ordinary effective 
potential~\cite{cw}, because the vector field propagator is 
transverse and annihilates mixed (scalar-vector) terms. 
Similarly, supersymmetric Landau gauge simplifies the computation
of the effective K\"ahler potential \cite{wgr,gru}, 
because the vector superfield propagator is 
`supertransverse' and annihilates mixed 
(chiral-vector) terms. 
So we will stick to this choice\footnote{We will not discuss the 
$\xi$ dependence of the one-loop K\"ahler potential. For such a 
study in renormalizable theories, where the $\xi$ dependence
affects finite terms, see \cite{gru} (and also \cite{pw} for 
the case $\xi$=1).}. 
The nice properties of supersymmetric Landau gauge are 
transferred to the component level, as they should. 
In particular, the mixed terms in (\ref{lpv}) become irrelevant
(in the constant $\hph$ background), so mixed $\phi V$ propagators 
are not generated and the $\phi\phi$ and $VV$ propagators found 
above are not modified. 
Indeed, the combinations of vector superfield components 
contained in (\ref{lpv}) are the same that appear in the 
gauge fixing lagrangian (last line of (\ref{lvv})), and 
those components become non-propagating for $\xi \rightarrow 0$.
As a cross-check, it is easy to verify explicitly that, for 
$\xi \rightarrow 0$, the mixed terms in (\ref{lpv}) are annihilated 
by the vector supermultiplet propagators\footnote{In more detail: 
$F G$ terms do not contribute because 
now $G$ has vanishing propagator; mixed terms of the type 
$\ph \, \dmu A_{\mu}$ are annihilated by the $A_{\mu}$ propagator 
as usual; similarly, the structure of propagators in the $(D,C)$ 
and $(\la,\chi)$ sectors is such that the $ \ph (D,C)$ and 
$\psi (\la,\chi)$ terms are annihilated, too.}.

\section{No corrections to the superpotential}

Our next computation is the $\hph$-dressed one-point function 
of $\hf$, at one-loop level. This gives us the term 
$\Delta w_i(\hph) \hf^i$ in the effective lagrangian (\ref{deltall}),
so we can verify whether or not the superpotential receives a one-loop 
correction $\Delta w$. The quantum bilinears proportional to
$\hf$, eqs. (\ref{lppf}), (\ref{lvvf}) and (\ref{lpvf}),
should be contracted with propagators, in order to close
the $\hf$ tadpole. In supersymmetric Landau gauge, the 
required propagators are absent for most of those terms. 
The only possible contributions to the $\hf$ tadpole
come from the third and fourth terms in (\ref{lppf}), which
can be closed with propagators. However, the corresponding 
contributions have equal magnitude and opposite sign, so
they cancel each other: 
\vspace{-0.3 cm}
\be
\begin{picture}(60,40)(0,17) 
\DashArrowLine(0,20)(30,20){3}\Vertex(30,20){1}
\DashArrowArcn(45,20)(15,180,0){3} 
\DashArrowArc(45,20)(15,180,0){3}
\Text(0,23)[bl]{$\hf$}
\Text(60,35)[l]{$\ph$}
\Text(60,5)[l]{$F$}
\end{picture}
\;\;\; + \;\;\;
\begin{picture}(60,40)(0,17) 
\DashArrowLine(0,20)(30,20){3}\Vertex(30,20){1}
\ArrowArcn(45,20)(15,180,0)
\ArrowArc(45,20)(15,180,0)
\Text(0,23)[bl]{$\hf$}
\Text(60,35)[l]{$\psi$}
\Text(60,5)[l]{$\psi$}
\end{picture}
\;\;\; = \;\;\; 0
\ee
\vspace{0.3 cm}\\
Thus the total $\hf$ tadpole vanishes, which we interpret as 
$\Delta w =0$. For generic $\xi$, the same result is obtained as 
a consequence of more complicated cancellations among several
component diagrams involving both chiral and vector supermultiplets
(we have checked this on specific examples). 
The absence of one-loop corrections to the superpotential in 
the general theory under study can be regarded as an explicit check 
of the well known non-renormalization theorem \cite{grs} (see also
\cite{sei,pr,wein}).
We recall that the literature contains some examples which violate 
the theorem at the one- or two-loop level, due to infrared effects 
associated to massless particles \cite{west,buchw,bcp}. We do not
find such violations, because the background field $\hph$ 
generates effective mass terms and thus acts as an effective 
infrared regulator in field space \cite{cw,pr}.
 
\section{Corrections to the K\"ahler potential}

We now move to the main computation, which is the 
$\hph$-dressed two-point function of $\hf$, at one-loop level
and vanishing external momenta, in supersymmetric Landau gauge. 
This will give us the term 
$\Delta K_{{\ov\imath} j}(\hphb,\hph) \hfb\,^{\ov\imath} \hf^j$ 
in the effective lagrangian (\ref{deltall}), so we will 
obtain $\Delta K$. The contribution of each 
diagram to the effective lagrangian is given explicitly,
and the integration in (Minkowski) momentum space, 
$\int d^4 p/(2\pi)^4$, is denoted by $\int_p$. Although the 
integrals have quadratic or logarithmic ultraviolet divergences, 
for the time being we do not select a specific regularization. In 
this respect, notice that no dangerous shifts in the loop momentum 
$p$ are needed, since the diagrams are evaluated at vanishing 
external momenta. Three classes of diagrams have to be considered:
they involve the contributions of $\phi$ multiplets only, 
$V$ multiplets only, or both.

\vspace{.3 cm}
{\em a) Pure $\phi$ loops.}  Using the first interaction term
in $\cl^{(\hf\hf)}_{\phi\phi}$, eq.~(\ref{lppff}), and the first 
and second ones (+h.c.) in $\cl^{(\hf)}_{\phi\phi}$, 
eq.~(\ref{lppf}), we obtain these contributions:
\vspace{-0.5 cm}
\bea
\begin{picture}(70,40)(0,10) 
\DashArrowLine(0,0)(35,0){3}\Vertex(35,0){1}
\DashArrowLine(35,0)(70,0){3}
\DashArrowArc(35,15)(15,-90,90){3}
\DashArrowArc(35,15)(15,90,270){3} 
\Text(0,3)[bl]{$\hf$}
\Text(70,3)[br]{$\hf$}
\Text(20,30)[tr]{$\ph$}
\Text(50,30)[tl]{$\ph$}
\end{picture}
& \!= \!&\!\! \hfb\,^{\ov\imath} {\hf^j}
\! \int_p \!\! i \,{\rm Tr}\! \left[\! \hk_{\ov\imath j}
\!\left(\! \hk p^2 - \hwb \hk^{-1 \, T} \hw \!\right)^{\!-1}\! \right]
\label{a1}
\\
\begin{picture}(70,40)(0,17) 
\DashArrowLine(0,20)(20,20){3}\Vertex(20,20){1}
\DashArrowLine(50,20)(70,20){3}\Vertex(50,20){1}
\DashArrowArc(35,20)(15,90,180){3}
\DashArrowArcn(35,20)(15,270,180){3} 
\DashArrowArc(35,20)(15,0,90){3}
\DashArrowArcn(35,20)(15,0,-90){3}
\Text(0,23)[bl]{$\hf$}
\Text(70,23)[br]{$\hf$}
\Text(20,35)[r]{$\ph$}
\Text(20,5)[r]{$\ph$}
\Text(50,35)[l]{$\ph$}
\Text(50,5)[l]{$\ph$}
\end{picture}
& \!= \!& \!\! -{1 \over 2} \hfb\,^{\ov\imath} {\hf^j}
\! \int_p \! \! i \,{\rm Tr} 
\!\left[
\!\left(\! \hk p^2 - \hwb \hk^{-1 \, T} \hw \right)^{\!-1} 
\!\! \hwb_{\ov\imath} 
\!\left(\! \hk^T p^2 - \hw \hk^{-1} \hwb \!\right)^{\!-1} 
\!\! \hw_j 
\!\right]
\label{a2}
\\
\begin{picture}(70,40)(0,17) 
\DashArrowLine(0,20)(20,20){3}\Vertex(20,20){1}
\DashArrowLine(50,20)(70,20){3}\Vertex(50,20){1}
\DashArrowArc(35,20)(15,90,180){3}
\DashArrowArc(35,20)(15,180,270){3} 
\DashArrowArc(35,20)(15,0,90){3}
\DashArrowArc(35,20)(15,-90,0){3}
\Text(0,23)[bl]{$\hf$}
\Text(70,23)[br]{$\hf$}
\Text(20,35)[r]{$\ph$}
\Text(20,5)[r]{$F$}
\Text(50,35)[l]{$\ph$}
\Text(50,5)[l]{$F$}
\end{picture}
&\! =\! &\!\! - \hfb\,^{\ov\imath} {\hf^j}
\! \int_p \!\! i p^2 \, {\rm Tr} 
\!\left[\! 
\left(\! \hk p^2 - \hwb \hk^{-1 \, T} \hw \!\right)^{\!-1} 
\!\! \hk_{\ov\imath} 
\!\left(\! \hk p^2 - \hwb \hk^{-1 \, T} \hw \!\right)^{\!-1} 
\!\! \hk_j 
\!\right]
\label{a3}
\\
\begin{picture}(70,40)(0,17) 
\DashArrowLine(0,20)(20,20){3}\Vertex(20,20){1}
\DashArrowLine(50,20)(70,20){3}\Vertex(50,20){1}
\DashArrowArc(35,20)(15,90,180){3}
\DashArrowArc(35,20)(15,180,270){3} 
\DashArrowArcn(35,20)(15,90,0){3}
\DashArrowArcn(35,20)(15,0,-90){3}
\Text(0,23)[bl]{$\hf$}
\Text(70,23)[br]{$\hf$}
\Text(20,35)[r]{$\ph$}
\Text(20,5)[r]{$F$}
\Text(50,35)[l]{$F$}
\Text(50,5)[l]{$\ph$}
\end{picture}
& \!=\! &\!\! - \hfb\,^{\ov\imath} {\hf^j}
\! \int_p \!\! i \,{\rm Tr} 
\!\left[
\!\left(\! \hk p^2 - \hwb \hk^{-1 \, T} \hw \!\right)^{\!-1} 
\!\! \hwb \hk^{-1 \, T} \hk^T_{\ov\imath} 
\!\left(\! \hk^T p^2 - \hw \hk^{-1} \hwb \!\right)^{\!-1} 
\!\! \hw \hk^{-1} \hk_j 
\!\right]
\nonumber
\\
& & 
\label{a4}
\\
\begin{picture}(70,40)(0,17) 
\DashArrowLine(0,20)(20,20){3}\Vertex(20,20){1}
\DashArrowLine(50,20)(70,20){3}\Vertex(50,20){1}
\DashArrowArc(35,20)(15,90,180){3}
\DashArrowArc(35,20)(15,180,270){3} 
\DashArrowArc(35,20)(15,0,90){3}
\DashArrowArcn(35,20)(15,0,-90){3}
\Text(0,23)[bl]{$\hf$}
\Text(70,23)[br]{$\hf$}
\Text(20,35)[r]{$\ph$}
\Text(20,5)[r]{$F$}
\Text(50,35)[l]{$\ph$}
\Text(50,5)[l]{$\ph$}
\end{picture}
&\! =\! & \!\! \hfb\,^{\ov\imath} {\hf^j}
\! \int_p \!\! i \, {\rm Tr} 
\! \left[ 
\! \left(\! \hk p^2 - \hwb \hk^{-1 \, T} \hw \!\right)^{\!-1} 
\!\! \hwb_{\ov\imath} 
\! \left(\! \hk^T p^2 - \hw \hk^{-1} \hwb \!\right)^{\!-1} 
\!\! \hw \hk^{-1} \hk_j 
\!\right]
\label{a5}
\\
\begin{picture}(70,40)(0,17) 
\DashArrowLine(0,20)(20,20){3}\Vertex(20,20){1}
\DashArrowLine(50,20)(70,20){3}\Vertex(50,20){1}
\DashArrowArc(35,20)(15,90,180){3}
\DashArrowArcn(35,20)(15,270,180){3} 
\DashArrowArc(35,20)(15,0,90){3}
\DashArrowArc(35,20)(15,-90,0){3}
\Text(0,23)[bl]{$\hf$}
\Text(70,23)[br]{$\hf$}
\Text(20,35)[r]{$\ph$}
\Text(20,5)[r]{$\ph$}
\Text(50,35)[l]{$\ph$}
\Text(50,5)[l]{$F$}
\end{picture}
&\! =\! & \!\! \hfb\,^{\ov\imath} {\hf^j}
\! \int_p \!\! i \, {\rm Tr} 
\! \left[ \!
\left(\! \hk p^2 - \hwb \hk^{-1 \, T} \hw \!\right)^{-1} 
\!\! \hk_{\ov\imath} 
\!\left(\! \hk p^2 - \hwb \hk^{-1 \, T} \hw \!\right)^{-1} 
\!\! \hwb \hk^{-1 \, T} \hw_j 
\! \right]
\label{a6}
\eea
\vspace{0.2 cm}\\
We recall that here $\hw$ and $\hk$ denote the {\em matrices}
defined in (\ref{hkhw}), and $\hw_j,\hk_j,\hk_{\ov\imath j},
\ldots$ denote derivatives of those matrices (i.e. third
derivatives of $w$, third and fourth derivatives of $K$).
Two additional diagrams can be built, using the third and 
fourth terms (+h.c.) in  $\cl^{(\hf)}_{\phi\phi}$, 
eq.~(\ref{lppf}). However, they cancel each other:
\vspace{-0.5 cm}
\be
\begin{picture}(70,40)(0,17) 
\DashArrowLine(0,20)(20,20){3}\Vertex(20,20){1}
\DashArrowLine(50,20)(70,20){3}\Vertex(50,20){1}
\DashArrowArcn(35,20)(15,180,90){3}
\DashArrowArc(35,20)(15,180,270){3} 
\DashArrowArcn(35,20)(15,90,0){3}
\DashArrowArc(35,20)(15,-90,0){3}
\Text(0,23)[bl]{$\hf$}
\Text(70,23)[br]{$\hf$}
\Text(20,35)[r]{$\ph$}
\Text(20,5)[r]{$F$}
\Text(50,35)[l]{$\ph$}
\Text(50,5)[l]{$F$}
\end{picture}
\;\;\; + \;\;\;
\begin{picture}(70,40)(0,17) 
\DashArrowLine(0,20)(20,20){3}\Vertex(20,20){1}
\DashArrowLine(50,20)(70,20){3}\Vertex(50,20){1}
\ArrowArcn(35,20)(15,180,90)
\ArrowArc(35,20)(15,180,270) 
\ArrowArcn(35,20)(15,90,0)
\ArrowArc(35,20)(15,-90,0)
\Text(0,23)[bl]{$\hf$}
\Text(70,23)[br]{$\hf$}
\Text(20,35)[r]{$\psi$}
\Text(20,5)[r]{$\psi$}
\Text(50,35)[l]{$\psi$}
\Text(50,5)[l]{$\psi$}
\end{picture}
\;\;\; = \;\;\; 0
\ee

\vspace{0.5 cm}
{\em b) Pure $V$ loops.} Using the interaction terms 
in  $\cl^{(\hf\hf)}_{VV}$, eq.~(\ref{lvvff}),
and  $\cl^{(\hf)}_{VV}$, eq.~(\ref{lvvf}), we obtain:
\vspace{-0.5 cm}
\bea
\begin{picture}(70,40)(0,10) 
\DashArrowLine(0,0)(35,0){3}\Vertex(35,0){1}
\DashArrowLine(35,0)(70,0){3}
\DashCArc(35,15)(15,-90,90){3}
\DashCArc(35,15)(15,90,270){3} 
\Text(0,3)[bl]{$\hf$}
\Text(70,3)[br]{$\hf$}
\Text(20,30)[tr]{$C$}
\Text(50,30)[tl]{$C$}
\end{picture}
& = & \hfb\,^{\ov\imath} {\hf^j}
\int_p   {i \over p^2} \, {\rm Tr}
\left[   \hx_{\ov\imath j} 
\left( \hr \, p^2 - \hx \right)^{-1} \right]
\label{b1}
\\
\begin{picture}(70,40)(0,17) 
\DashArrowLine(0,20)(20,20){3}\Vertex(20,20){1}
\DashArrowLine(50,20)(70,20){3}\Vertex(50,20){1}
\ArrowArc(35,20)(15,90,180)
\ArrowArcn(35,20)(15,270,180) 
\ArrowArc(35,20)(15,0,90)
\ArrowArcn(35,20)(15,0,-90)
\Text(0,23)[bl]{$\hf$}
\Text(70,23)[br]{$\hf$}
\Text(20,35)[r]{$\lambda$}
\Text(20,5)[r]{$\lambda$}
\Text(50,35)[l]{$\lambda$}
\Text(50,5)[l]{$\lambda$}
\end{picture}
& = & \hfb\,^{\ov\imath} {\hf^j}
\int_p \! i p^2 \, {\rm Tr} \left[  
\hr_{\ov\imath} \left( \hr  \, p^2 - \hx \right)^{-1} 
\hr_j \left( \hr  \, p^2 - \hx \right)^{-1} 
\right]
\label{b2}
\\
\begin{picture}(70,40)(0,17) 
\DashArrowLine(0,20)(20,20){3}\Vertex(20,20){1}
\DashArrowLine(50,20)(70,20){3}\Vertex(50,20){1}
\ArrowArcn(35,20)(15,180,90)
\ArrowArc(35,20)(15,180,270) 
\ArrowArcn(35,20)(15,90,0)
\ArrowArc(35,20)(15,-90,0)
\Text(0,23)[bl]{$\hf$}
\Text(70,23)[br]{$\hf$}
\Text(20,35)[r]{$\chi$}
\Text(20,5)[r]{$\chi$}
\Text(50,35)[l]{$\chi$}
\Text(50,5)[l]{$\chi$}
\end{picture}
& = &  \hfb\,^{\ov\imath} {\hf^j}
\int_p  {i \over p^2} \,  {\rm Tr} \left[ 
\hx_{\ov\imath} \left( \hr  \, p^2 - \hx \right)^{-1} 
\hx_j \left( \hr  \, p^2 - \hx \right)^{-1} 
\right]
\label{b3}
\\
\begin{picture}(70,40)(0,17) 
\DashArrowLine(0,20)(20,20){3}\Vertex(20,20){1}
\DashArrowLine(50,20)(70,20){3}\Vertex(50,20){1}
\ArrowArc(35,20)(15,90,180)
\ArrowArcn(35,20)(15,270,180) 
\ArrowArcn(35,20)(15,90,0)
\ArrowArc(35,20)(15,-90,0)
\Text(0,23)[bl]{$\hf$}
\Text(70,23)[br]{$\hf$}
\Text(20,35)[r]{$\lambda$}
\Text(20,5)[r]{$\lambda$}
\Text(50,35)[l]{$\chi$}
\Text(50,5)[l]{$\chi$}
\end{picture}
& = &  - \hfb\,^{\ov\imath} {\hf^j}
\int_p\! i \, {\rm Tr} \left[ 
\hx_{\ov\imath} \left( \hr  \, p^2 - \hx \right)^{-1} 
\hr_j \left( \hr  \, p^2 - \hx \right)^{-1} 
\right]
\label{b4}
\\
\begin{picture}(70,40)(0,17) 
\DashArrowLine(0,20)(20,20){3}\Vertex(20,20){1}
\DashArrowLine(50,20)(70,20){3}\Vertex(50,20){1}
\ArrowArcn(35,20)(15,180,90)
\ArrowArc(35,20)(15,180,270) 
\ArrowArc(35,20)(15,0,90)
\ArrowArcn(35,20)(15,0,-90)
\Text(0,23)[bl]{$\hf$}
\Text(70,23)[br]{$\hf$}
\Text(20,35)[r]{$\chi$}
\Text(20,5)[r]{$\chi$}
\Text(50,35)[l]{$\lambda$}
\Text(50,5)[l]{$\lambda$}
\end{picture}
& = & - \hfb\,^{\ov\imath} {\hf^j}
\int_p\! i \, {\rm Tr} \left[ 
\hr_{\ov\imath} \left( \hr \, p^2 - \hx \right)^{-1} 
\hx_j \left( \hr  \, p^2 - \hx \right)^{-1} 
\right]
\label{b5}
\eea
\vspace{0.2 cm}\\
In the above expressions, $\hx_j,\hr_j,\hr_{\ov\imath j},\ldots$
denote derivatives of the matrices $\hx$ and $\hr$ defined in
(\ref{hrhx}). Also, the trace operation does not include the 
trace in spinor space: the latter has already been performed.

\vspace{0.5 cm}
{\em c) Mixed $\phi$-$V$ loops.} Two non-vanishing diagrams can
be built using the interaction terms in $\cl^{(\hf)}_{\phi V}$, 
eq.~(\ref{lpvf}), but they cancel each other: 
\vspace{-0.5 cm}
\be
\begin{picture}(70,40)(0,17) 
\DashArrowLine(0,20)(20,20){3}\Vertex(20,20){1}
\DashArrowLine(50,20)(70,20){3}\Vertex(50,20){1}
\DashArrowArcn(35,20)(15,180,90){3}
\DashArrowArcn(35,20)(15,90,0){3}
\DashCArc(35,20)(15,180,0){3}
\Text(0,23)[bl]{$\hf$}
\Text(70,23)[br]{$\hf$}
\Text(20,35)[]{$F$}
\Text(20,5)[r]{$C$}
\Text(50,35)[l]{$F$}
\Text(50,5)[l]{$C$}
\end{picture}
\;\;\; + \;\;\;
\begin{picture}(70,40)(0,17) 
\DashArrowLine(0,20)(20,20){3}\Vertex(20,20){1}
\DashArrowLine(50,20)(70,20){3}\Vertex(50,20){1}
\ArrowArcn(35,20)(15,180,90)
\ArrowArc(35,20)(15,180,270) 
\ArrowArcn(35,20)(15,90,0)
\ArrowArc(35,20)(15,-90,0)
\Text(0,23)[bl]{$\hf$}
\Text(70,23)[br]{$\hf$}
\Text(20,35)[r]{$\psi$}
\Text(20,5)[r]{$\chi$}
\Text(50,35)[l]{$\psi$}
\Text(50,5)[l]{$\chi$}
\end{picture}
\;\;\; = \;\;\; 0
\ee
\vspace{0.2 cm}\\
This cancellation is a further effect of supersymmetric 
Landau gauge at the component level. In superspace, the 
vanishing of mixed $\phi$-$V$ loops should automatically
follow from mixed $\phi$-$V$ vertices being annihilated by the 
supertransverse $V$ propagator, as in the renormalizable
case \cite{gru}.
\vspace{0.3 cm}
 
Now we have to sum the diagrams above and express the coefficient 
of $\hfb\,^{\ov\imath} {\hf^j}$ as a second derivative 
(with respect to $\hphb\,^{\ov\imath}$ and $\hph^j$).
After some manipulations, the results for the $\phi$ and $V$ 
sectors can be cast in the required form:
\vspace{-0.5 cm}
\bea
\label{resp}
\begin{picture}(70,40)(0,17) 
\DashArrowLine(0,20)(20,20){3}\Vertex(20,20){1}
\DashArrowLine(50,20)(70,20){3}\Vertex(50,20){1}
\CArc(35,20)(15,0,360)
\Text(35,20)[]{\large $\phi$}
\Text(0,23)[bl]{$\hf$}
\Text(70,23)[br]{$\hf$}
\end{picture}
& = &  \hfb\,^{\ov\imath} {\hf^j}
\int_p {i \over p^2} \left[
{1 \over 2} \, {\rm Tr} \log \hk + {1 \over 2} 
\, {\rm Tr} \log 
\left( \hk p^2 - \hwb \hk^{-1 \; T} \hw \right) 
\right]_{\ov\imath j}
\\
\label{resv}
\begin{picture}(70,40)(0,17) 
\DashArrowLine(0,20)(20,20){3}\Vertex(20,20){1}
\DashArrowLine(50,20)(70,20){3}\Vertex(50,20){1}
\CArc(35,20)(15,0,360)
\Text(35,20)[]{\large $V$}
\Text(0,23)[bl]{$\hf$}
\Text(70,23)[br]{$\hf$}
\end{picture}
& = &  \hfb\,^{\ov\imath} {\hf^j}
\int_p {i \over p^2} \left[ -
{\rm Tr} \log \left(\hr \, p^2 - \hx \right) 
\right]_{\ov\imath j}
\eea

\vspace{0.2 cm}
Finally, from the comparison of eqs.~(\ref{resp}) and (\ref{resv}) with
eq.~(\ref{deltall}), we can read off the one-loop correction to the 
K\"ahler potential:
\be
\label{res}
\Delta K(\hphb,\hph)   =  
\int_p {i \over p^2} \left[
{1 \over 2}\, {\rm Tr} \log \hk + {1 \over 2}\, {\rm Tr} \log 
\left( \hk p^2 - \hwb \hk^{-1 \; T} \hw \right) 
- {\rm Tr} \log \left(\hr \, p^2 - \hx \right) 
\right] \, .
\ee
We can go one step further and perform the momentum integration.
If we do a Wick rotation and regulate the momentum integral  
with a simple ultraviolet cutoff $\Lambda$, the result 
reads:
\bea
\label{resbis} 
\Delta K(\hphb,\hph) & = &  {\lsq \over 16 \pi^2} 
\left[ \, \log {\rm det} \hk - \log {\rm det} \hr \, \right]
\nonumber \\
& & \nonumber \\
& - & 
{1 \over 32 \pi^2} \left[
{\rm Tr} \left( 
\cm^2_{\phi} \left( \log {\cm^2_{\phi} \over \lsq} \, -1 \right) \right)
- 2 \, {\rm Tr} \left(
\cm^2_V \left( \log {\cm^2_V \over \lsq} \, -1 \right) \right)
\right]
\eea
where $\cm^2_{\phi}$ and $\cm^2_V$ are field dependent
mass matrices in the chiral and vector sectors:
\be
\label{mm}
\cm^2_{\phi} \equiv \hk^{-1/ 2} \,\hwb \, \hk^{-1 \; T} \, \hw \,\hk^{-1/ 2}
\;\; , \;\;\;\; 
\cm^2_V \equiv \hr^{-1/2} \, \hx \, \hr^{-1/2} \, .
\ee
We recall that, in the above expressions, all the dependence 
on $(\hphb,\hph)$ is contained in the matrices $\hk,\hw,\hr,\hx$ 
defined in (\ref{hkhw}) and (\ref{hrhx}). The functional dependence
of $\Delta K$ is what we were looking for. Indeed, if we recall that 
eq.~(\ref{deltall}) was derived from eq.~(\ref{deltal}), it is clear
that we can go back to superspace and replace the arguments 
$(\hphb,\hph)$ of $\Delta K$ with general superfields 
$(\hphib,\hphi)$, or even with $(\hphib,e^{2 \hv} \hphi)$ 
in the case of a background gauge invariant quantization.
So eq.~(\ref{resbis}) is our final result: it gives the full 
one-loop correction to the K\"ahler potential in a closed form,
for the general theory under study. 
The first line of eq.~(\ref{resbis}) contains quadratically divergent
contributions, whereas the second line contains logarithmically 
divergent and finite contributions. 
If we evaluate the momentum integral in 
$d\!=\!4 - 2 \epsilon$ dimensions 
($\int_p=\mu^{2\epsilon} \int d^dp/(2\pi)^d$) instead of using
a momentum cutoff in $d\!=\!4$,  
the first line of eq.~(\ref{resbis}) should be omitted, and  
the replacement $\log \lsq \rightarrow 1/\epsilon \,+ 1 
- \gamma + \log(4\pi\mu^2)$ should be made in the second 
line\footnote{ Notice that this regularization corresponds to 
dimensional reduction \cite{siegel}, since the spinor algebra 
has been performed in $d\!=\!4$.}.

In the special case of renormalizable theories, the superpotential 
is at most cubic and the metric is canonical in both the chiral and 
vector sectors, so 
$\hk_{\ov\imath j}=\delta_{\ov\imath j}={\rm const.}$, 
$\hw_{ij}=m_{ij}+h_{ijk}\hph^k$, $\hr_{ab}=\delta_{ab}/g_a^2
={\rm const.}$, $\hx_{ab}=\hphb \{ T_a, T_b \} \hph$.
In this limit, the first line of (\ref{resbis}) becomes irrelevant 
and the second line reproduces the result derived in \cite{gru}, 
if $g_a^2$ is identified with $2 g^2$. Incidentally, 
we recall that the result of \cite{gru} was obtained by 
computing two superdiagrams,
i.e. one in each sector, after resumming $\hphi$ insertions. 
Our component field approach is not much more involved. 
Using $\hph$-dressed propagators amounts to resumming
$\hph$ insertions, and our result for $\Delta K$ originates 
from only three $\hfb\hf$ component diagrams in the 
renormalizable case, i.e. (\ref{a2}) in the chiral 
sector and (\ref{b1}), (\ref{b3}) in the vector sector.

In the case of non-renormalizable theories, the quadratically 
divergent contributions in the first line of (\ref{resbis})
agree with the results of ref.~\cite{grk}, obtained with 
superfield methods. In our approach, those contributions
can be traced back to the component diagrams (\ref{a1}), 
(\ref{a3}) and (\ref{b2}). We have also tried to make a 
comparison with ref.~\cite{maryk}, in which detailed computations 
of the one-loop bosonic effective action in general supergravity 
theories with diagonal gauge kinetic function were presented,
and the flat limit was also considered. 
Only the divergent contributions were evaluated, and part
of them was interpreted as a correction to the K\"ahler
potential. A component field approach different from ours
was used. 
Here we have insisted on preserving supersymmetry and supersymmetric
background gauge invariance. In \cite{maryk} the Wess-Zumino
gauge was used, and special emphasis was given to ordinary background 
gauge invariance and scalar field reparametrization covariance.
We recall that the Wess-Zumino gauge generally leads to a loss of 
manifest supersymmetry, since vector
supermultiplets are integrated out in a non-supersymmetric way. 
To compensate for this, a special $R_{\xi}$-type gauge fixing 
for the component vector fields was introduced in \cite{maryk}, 
with $\xi$=1.
This particular prescription was argued to restore supersymmetry,
since the anomalous dimensions of component scalar fields were 
found to coincide with the supersymmetric ones of the associated
chiral superfields, in the flat limit. Although this coincidence
may be partly accidental\footnote{For 
instance, extending that coincidence to the fermionic components of 
chiral superfields would require some additional modification. 
As a further example of one-loop computations in the Wess-Zumino gauge,
we may recall the component approach employed in \cite{bar}
to study the divergences in a general renormalizable
theory. In this case ordinary Landau gauge was used for the component 
vector fields, and the scalar field anomalous dimensions did 
not coincide with the chiral superfield ones. The latter were 
reconstructed from the renormalization of superpotential parameters.}, 
the divergent part of the one-loop K\"ahler potential reconstructed 
in \cite{maryk} seems to agree with ours. 
Strictly speaking, a slight difference can be found, for 
another reason. Indeed, the derivatives in $\hw_{ij}$ are 
reparametrization covariant ones in the formulae of 
\cite{maryk} (see also \cite{bpr}). 
However, this apparent discrepancy is not a physical one: it 
depends on the way the background-quantum splitting of chiral 
supermultiplets is performed\footnote{We recall that the 
perturbative computations and the resulting effective action 
depend on both that choice and other ones, such as the  
background-quantum splitting of vector supermultiplets and 
the choice of the gauge fixing function (or parameter). 
All such ambiguities are expected to disappear at the 
level of the physical S-matrix.}.
If desired, our result could be made reparametrization covariant 
{\it a posteriori}, e.g. by reinterpreting the derivatives in 
$\hw_{ij}$ as covariant ones and promoting $(T_a \hph)^i$ to a 
more general holomorphic Killing vector $v_a^i(\hph)$. 
This could perhaps be confirmed by a supersymmetric normal 
coordinate expansion \cite{ag,grk}, which however goes beyond 
the scope of the present paper.

\section{A consistency check}

We conclude by presenting an alternative derivation of $\Delta K$.
This derivation is based on background gauge invariance, and the
agreement of the final result with the one found above 
provides an interesting consistency check. We recall that,
although we have presented our general framework in a background 
gauge invariant way, the latter property has not been  
exploited in the previous section, where the functional form of 
$\Delta K$ has been computed by using a background with vanishing $\hv$. 
Strictly speaking, what we obtained was $\Delta K(\hphib,\hphi)$.
If the background-quantum splitting of vector 
superfields and the gauge fixing term are chosen as in (\ref{vsplit})
and (\ref{gfix}), the result should be automatically promoted to 
$\Delta K (\hphib,e^{2 \hv} \hphi)$. We want to check this 
explicitly, so we switch on a non-vanishing $\hv$, which we take 
to consist of a constant $\hd$ field, as in eq.~(\ref{backv}).
At the same time, we take the background superfield $\hphi$ in the
form (\ref{backp}), with constant $\hph$ but vanishing $\hf$.
Then we compute the terms linear in $\hd$ in the one-loop 
effective lagrangian (or potential), which eq. (\ref{deltall})
predicts to have the form $\hd^a [\Delta K_j(\hphb,\hph)(T_a \hph)^j 
+\Delta \kappa_a]$. Thus $\Delta K$ can be reconstructed and compared 
to the previous result, and $\Delta \kappa_a$ can be identified 
as well.

In order to check all this, we first have to expand the lagrangian
in the new background and find the terms bilinear in quantum 
fields ($\cl_{\rm bil}$) that have a linear dependence on $\hd$. 
We omit the full list because only a few among such terms give 
a non-vanishing $\hd$ tadpole, in supersymmetric Landau gauge. 
For instance, terms that couple $\hd$ to a mixed $\phi$-$V$ 
bilinear cannot contribute, due to the absence of mixed propagators.
Also, terms that couple $\hd$ to ghost bilinears are $\hph$ 
independent and could at most contribute to the 
Fayet-Iliopoulos term, but the actual contribution is zero 
because ghosts belong to the adjoint representation,
which is vector-like. Special care is needed to study the effect
of the gauge fixing lagrangian (\ref{gfix}), because it
generates terms that couple $\hd^a$ to the components of quantum
supermultiplets $V^b,V^c$ with strength $\sim c_{bc}^{\;\;\;a}/\xi$.
Since this coefficient is divergent in the limit $\xi \rightarrow 0$,
a small non-vanishing $\xi$ should be kept in intermediate steps,
and the mixed $\phi$-$V$ terms in $\cl^{(0)}_{\phi V}$,
eq.~(\ref{lpv}), should be taken into account. 
When pure $V^b$-$V^c$ propagators are used to close the $\hd$ tadpole, 
the integrand is zero because the propagators are $bc$-symmetric 
whereas the structure constants are antisymmetric\footnote{
A similar mechanism also kills other terms in the lagrangian.}.
If mixed $\phi$-$V$ insertions are used, at least two of them
are needed to close the loop. However, since they pick up
only the $\xi$ dependent parts of the adjacent $V$ 
propagators, the singular $1/\xi$ factor is multiplied 
by a factor at least $\co(\xi^2)$, so the result is again zero
in the limit $\xi \rightarrow 0$. 
After completing the inspection of these and other terms, we 
find that the only terms that can give a non-vanishing 
$\hd$ tadpole are\footnote{We remark that the  
background-quantum splitting of the vector superfield,
eq.~(\ref{vsplit}), plays a crucial role here. The simple
splitting $\hv \rightarrow \hv + V$ would not give
the desired result, for a non-abelian gauge group.}:
\bea
\label{lppd}
\cl_{\phi \phi}^{(\hd)} & = & \hd^a 
\left( \hk_{\ov\imath k} (T_a)^k_{\; j}
+ \hk_{\ov\imath k j} (T_a \hph)^k \right)
\phb^{\ov\imath} \ph^j 
\\
\label{lvvd}
\cl_{V V}^{(\hd)} & = & \hd^a \left[ 
{1 \over 2} (\hx_{bc})_j (T_a \hph)^j C^b C^c 
+{i \over 2} \hff_{bd} c_{ac}^{\;\;\; d} (D^b C^c - \la^b \chi^c)
\right] + {\rm h.c.}
\eea
The contributions to the $\hph$-dressed $\hd$ tadpole are,
in matrix notation:
\vspace{-0.5 cm}
\bea
\label{dpp}
\begin{picture}(50,40)(0,17) 
\DashLine(0,20)(20,20){3}\Vertex(20,20){1}
\DashArrowArc(35,20)(15,0,180){3} 
\DashArrowArc(35,20)(15,180,0){3}
\Text(0,23)[bl]{$\hd$}
\Text(50,35)[l]{$\ph$}
\Text(50,5)[l]{$\ph$}
\end{picture}
& \; = & 
\hd^a \int_p \! i\, {\rm Tr} \left[
\left( \hk T_a + \hk_j (T_a \hph)^j \right)
\left( \hk p^2 - \hwb \hk^{-1 \; T} \hw \right)^{-1} \right]
\\
\label{dcc}
\begin{picture}(50,40)(0,17) 
\DashLine(0,20)(20,20){3}\Vertex(20,20){1}
\DashCArc(35,20)(15,0,180){3} 
\DashCArc(35,20)(15,180,0){3}
\Text(0,23)[bl]{$\hd$}
\Text(50,35)[l]{$C$}
\Text(50,5)[l]{$C$}
\end{picture}
& \; = & 
- {1\over 2 } \hd^a \int_p {i\over p^2} \, {\rm Tr} \left[ 
\left( \hr \, p^2 - \hx \right)^{-1}  
\hx_j \right] (T_a \hph)^j + {\rm h.c.}
\\
\label{ddc}
\begin{picture}(50,40)(0,17) 
\DashLine(0,20)(20,20){3}\Vertex(20,20){1}
\DashCArc(35,20)(15,0,180){3} 
\DashCArc(35,20)(15,180,0){3}
\Text(0,23)[bl]{$\hd$}
\Text(50,35)[l]{$D$}
\Text(50,5)[l]{$C$}
\end{picture}
& \; = & 
{1\over 2} \hd^a \int_p \! i \, {\rm Tr} \left[ 
\left( \hr \, p^2 - \hx \right)^{-1}  
\hr_j \right] (T_a \hph)^j + {\rm h.c.}
\eea
\vspace{-0.3 cm}
\be
\label{dchla}
\begin{picture}(130,40)(0,17) 
\DashLine(0,20)(20,20){3}\Vertex(20,20){1}
\ArrowArc(35,20)(15,0,180) 
\ArrowArcn(35,20)(15,0,180)
\Text(0,23)[bl]{$\hd$}
\Text(50,35)[l]{$\chi$}
\Text(50,5)[l]{$\lambda$}
\Text(65,20)[]{$+$}
\DashLine(80,20)(100,20){3}\Vertex(100,20){1}
\ArrowArc(115,20)(15,180,0) 
\ArrowArcn(115,20)(15,180,0)
\Text(80,23)[bl]{$\hd$}
\Text(130,35)[l]{$\chi$}
\Text(130,5)[l]{$\lambda$}
\end{picture}
 \;\;  =  \; 
- \hd^a \int_p \! i \, {\rm Tr} \left[ 
\left( \hr \, p^2 - \hx \right)^{-1}  
\hr_j \right] (T_a \hph)^j + {\rm h.c.}
\ee
\vspace{0.2 cm}\\
On writing the expressions in (\ref{ddc}) and (\ref{dchla}), 
we have used the identity (\ref{fid}). The expression 
in (\ref{dpp}) can be put in a more suitable form by
using the relation $\hw T_a + T_a^T \hw = -\hw_j (T_a \hph)^j$
for the matrix $\hw$, which follows from the identity (\ref{wid}).
After some manipulations, the coefficient of $\hd^a$ in the 
$\phi$ and $V$ sectors can be cast in the form prescribed
by eq.~(\ref{deltall}):
\vspace{-0.5 cm}
\bea
\label{f4abis}
\begin{picture}(50,40)(0,17) 
\DashLine(0,20)(20,20){3}\Vertex(20,20){1}
\CArc(35,20)(15,0,360)
\Text(0,23)[bl]{$\hd$}
\Text(35,20)[]{\large $\phi$}
\end{picture}
& \! =\! &  
 \hd^a (T_a \hph)^j \!\! \int_p  {i \over p^2}\! \left[
{1 \over 2}  {\rm Tr} \log \hk \! +\! {1 \over 2} {\rm Tr} \log 
\! \left(\! \hk p^2 \! -\! \hwb \hk^{-1 \, T} \hw\! \right)\! 
\right]_j \!\! + \!\hd^a \,{\rm Tr}\, T_a \! \int_p {i \over p^2} 
\phantom{xxxxxx}
\\
\label{f4abc}
\begin{picture}(50,40)(0,17) 
\DashLine(0,20)(20,20){3}\Vertex(20,20){1}
\CArc(35,20)(15,0,360)
\Text(0,23)[bl]{$\hd$}
\Text(35,20)[]{\large $V$}
\end{picture}
&\! =\! &  \hd^a (T_a \hph)^j \! \int_p  {i \over p^2}\!
\left[ -{\rm Tr} \log \left( \hr \, p^2 - \hx \right) 
\right]_j
\eea

\vspace{0.2 cm} 
From the comparison of eqs.~(\ref{f4abis}) and (\ref{f4abc})
with eq.~(\ref{deltall}), we can easily identify $\Delta K$
and $\Delta\kappa_a$. The correction to the Fayet-Iliopoulos
coefficients is visible in the last term of eq.~(\ref{f4abis}),
and confirms a well known result \cite{fisc}:
\be
\Delta \kappa_a = {\rm Tr}\, T_a \int_p {i \over p^2}
= {\lsq \over 16 \pi^2}   {\rm Tr}\, T_a \, .
\ee
The correction to the K\"ahler potential can be read off
from the first part of eq.~(\ref{f4abis}) and from
eq.~(\ref{f4abc}): the expression of $\Delta K$
is identical to that obtained in eq.~(\ref{res})
using a different method. 
This completes our consistency check.

\end{document}